\documentclass[aps,prb,preprint,showpacs,superscriptaddress]{revtex4}
\usepackage{bm,amsmath,amssymb,latexsym,mathrsfs,enumerate}
\usepackage[mathcal]{euscript}
\usepackage{graphicx, subfigure}
\newcommand{\figwidth}{0.47\textwidth}

\begin{document}

\title{Chirality tunneling and quantum dynamics for domain walls in
mesoscopic ferromagnets}

\author{E.~G. Galkina}
\affiliation{Frontier Research System, The Institute of Physical
and Chemical Research (RIKEN), Wako-shi, Saitama, 351-0198, Japan}
\affiliation{Institute of Physics,  03028, Kiev, Ukraine }
\author{ B.~A. Ivanov}
\email{bivanov@i.com.ua} \affiliation{Frontier Research System,
The Institute of Physical and Chemical Research (RIKEN), Wako-shi,
Saitama, 351-0198, Japan} \affiliation{Institute of Magnetism,
03142 Kiev, Ukraine}
 \author{Sergey Savel'ev}
\affiliation{Frontier Research System, The Institute of Physical
and Chemical Research (RIKEN), Wako-shi, Saitama, 351-0198, Japan}
\affiliation{Department of Physics, Loughborough University,
Leicestershire LE11 3TU, United Kingdom}
 \author{Franco Nori}
 \affiliation{Frontier Research System, The Institute of Physical
and Chemical Research (RIKEN), Wako-shi, Saitama, 351-0198, Japan}
\affiliation{Department of Physics, Center for Theoretical
Physics, Applied Physics Program, Center for the Study of Complex
Systems, University of Michigan, Ann Arbor, MI 48109-1040, USA}

\date{\today}

\begin{abstract}

{We studied the quantum dynamics of ferromagnetic domain walls
(topological kink-type solitons) in one dimensional ferromagnetic
spin chains. We show that the tunneling probability does not
depend on the number of spins in a domain wall; thus, this
probability can be large even for a domain wall containing a large
number of spins. We also predict that there is a strong interplay
between the tunneling of a wall from one lattice site to another
(tunneling of the kink coordinate) and the tunneling of the kink
topological charge (so-called chirality). Both of these elementary
processes are suppressed for kinks in one-dimensional ferromagnets
with half-integer spin. The dispersion law (i.e., the domain wall
energy versus momentum) is essentially different for chains with
either integer or half-integer spins. The predicted quantum
effects could be observed for mesoscopic magnetic structures,
e.g., chains of magnetic clusters, large-spin molecules, or
nanosize magnetic dots.}

\end{abstract}

\pacs{75.10.Jm,  
03.75.Kk}



\maketitle

\section{Introduction}

Domain walls play an important role in the physics of magnets. For
macroscopic bulk magnetic samples, domain walls, being extended
classical objects, determine the demagnetizing processes, see,
e.g., Refs.~\onlinecite{MalozSlon,HubertBook,Bar-springer}. For
one-dimensional magnets, domain walls (kink-type solitons) play a
different role: they are nonlinear excitations responsible for the
destruction of long-range order. Thus, domain walls should be
taken into account together with the linear excitations
(magnons).\cite{MikStainer,BarIvKhalat} Quantum properties are
inherent to kinks in one-dimensional magnets (spin chains) with
small spin values, like $S =1/2$ or $S =1 $  and high
anisotropy.\cite{MikStainer}

Classical solitons in one-dimensional Heisenberg ferromagnets have
been investigated in detail. For continuum media, their dynamical
properties are determined by the Landau-Lifshitz equation for the
magnetization vector $\mathbf{m}(x, t)$, where $\mathbf{m}^2=1$,
see e.g., Refs. \onlinecite{MikStainer,Bar-springer,Kosevich+All}.
For such systems, kink-type solitons can be treated as classical
particle-like objects. However, kinks are extended objects, and
for spin chains with low anisotropy $K \ll J$ ($J$ is the exchange
integral and $K$ the anisotropy constant), a kink involves a large
number of spins $N_{\mathrm{kink}} \sim S\sqrt{J/K} \gg 1$. For
this reason, domain walls for low-anisotropy magnets should be
formally considered as a mesoscopic, rather than a microscopic,
object. Thus, it is not obvious whether or not the effects of
quantum coherence are essential for domain wall dynamics in
mesoscopic ferromagnets.

Artificial quasi-1D mesoscopic materials (including chains of
small magnetic elements, such as small magnetic particles of
nanometer size (magnetic dots), patterned magnetic films, magnetic
clusters and high-spin molecules) are of great
importance,\cite{Wernsd,Skomski,CowWelSci00-Computers,savel,otani,otani1,otani2,otani3}
and are promising elements for
computers.\cite{CowWelSci00-Computers} These materials often
manifest unique physical properties that are absent in bulk
samples, for instance, macroscopic quantum coherence and quantum
tunneling, see e.g., Refs.~\onlinecite{Wernsd,Skomski}. These
quantum properties of small magnetic systems could be potentially
useful for quantum computing.\cite{QuanComp}

Quantum coherence can occur when states with the same  energy are
separated by a small potential energy barrier. Kink-type solitons
can demonstrate a rich variety of different quantum effects.
Indeed solitons are particle-like objects and their quantum
dynamics include the tunneling of the kink coordinates through a
potential barrier, separating equivalent positions in 1D chains.
These tunneling effects are coherent, and, as for electrons in
crystals, they lead to the formation of band spectra.

Another type of tunneling involves the domain wall chirality.
Namely, a domain wall is characterized by the deviation of the
magnetization from the easy axis. The corresponding value of the
total spin $\mathbf{S}_{\perp}$ is quite large, i.e.,
$|\mathbf{S}_{\perp}|$ is of the order of $SN_{\mathrm{kink}} \gg
1$. For biaxial magnets, the domain wall state has two-fold
degeneracy along the $\mathbf{S}_{\perp}$ direction.
From a mathematical point of view, the sign of the quantity
$\mathbf{S}_{\perp}$ corresponds to the value of topological
charge or chirality $\chi = \pm 1$. Thus, the kink structure is
doubly-degenerate over the  sign of this topological charge,
implying the possibility of a quantum coherent superposition of
two states with different chirality.

The tunneling of topological charges has previously been discussed
for different topological solitons (kinks, vortices, and
disclinations) in antiferromagnets; for a review
see.\cite{IvanovFNT05} It is worth noting here that the static
distribution of the corresponding order parameters, the normalized
magnetization vector $\mathbf{m}$  for ferromagnets
($\mathbf{m}^2=1$)  and  sublattice magnetization vector
$\mathbf{l}$ ($\mathbf{l}^2=1$) for antiferromagnets, are all
identical. For antiferromagnetic spin chains, the rate of the
chirality tunneling process appears to be unexpectedly high,
because the tunneling exponent is of the order of the atomic spin
$S$, being independent on the number of spins $N_{\mathrm{kink}}
\gg 1$ within the kink.\cite{IvKol} Thus, quantum effects could be
essential not only for literally 1D objects like spin chains, but
also for mesoscopic antiferromagnetic samples, like thin
antiferromagnetic wires.\cite{IvKolKir98}

The dynamic properties of ferromagnets are described by the
Landau-Lifshitz equation for the vector $\mathbf{m}$ with no
inertial term. This is in strong contrast with the inertial
dynamics of antiferromagnets, described by the so-called
sigma-model equation for the sublattice magnetization vector
$\mathbf{l}$, e.g., Refs.~\onlinecite{MikStainer},
\onlinecite{Bar-springer}. It might look a bit paradoxical that
the quantum properties of ferromagnets are much more complicated
than for both quantum antiferromagnets and quantum Josephson
junctions~\cite{mqt,mqt1,mqt2}. The reason is because within the
sigma-model approach the dynamics of the vector $\mathbf{l}$ is
similar to that of the usual inertial dynamics of a particle
(strictly speaking, the dynamics of a particle along the surface
of the sphere $\mathbf{l}^2=1$); whereas the Landau-Lifshitz
Lagrangian contains a complicated Dirac-monopole term with
non-trivial topological properties (Berry phase), see, e.g.,
Ref.~\onlinecite{FRA91}. This circumstance leads to a number of
subtle and intriguing effects, e.g., the suppression of tunneling
transitions due to the interference of the instanton
trajectories.\cite{los-div-gri92,del-hen92}

An example of strikingly different quantum dynamics of
ferromagnets and antiferromagnets is the tunneling chirality. For
kinks in antiferromagnets, chirality tunneling is not correlated
with the translational motion of kinks.\cite{IvKol,IvKolKir98} In
contrast, some results known in the literature imply that for
ferromagnets the situation can be different.
Ref.~\onlinecite{BraunLoss96} noted that unmovable kinks with
different values of the chirality must have different values of
the momentum. Ref.~\onlinecite{TakagiTatara96} pointed out that
the chirality tunneling rate grows with the intensity of the kink
spatial pinning. Ref.~\onlinecite{ShibTakagi00} showed that for a
free (i.e., with no pinning) domain wall in a ferromagnet, the
momentum of which is conserved, the tunneling of the chirality is
prohibited.

In this article we develop a consistent quantum theory of domain
walls in one-dimensional ferromagnets, with a complete description
of all possible coherent quantum effects. We show that the tunneling
of the chirality can be described as a tunneling in momentum space.
This is closely connected with the tunneling of the kink
coordinates; indeed, all these tunneling processes are naturally
described within phase plane $(X,P)$.

The article is organized as follows. In the next section \ref{2},
we introduce the Hamilton variables: kink coordinate $X$ and kink
momentum $P$. In this section \ref{2} we show how to consistently
define the chirality of a kink via the value of kink momentum. A
Hamiltonian approach, valid for describing both quantum effects,
tunneling of kink coordinates and kink chirality, will be
developed in the same Section \ref{2}. Then, in Section \ref{3},
specific tunneling effects will be analyzed based on this
approach. Kink dispersion relations will also be derived there.

\section{Hamilton description of kink dynamics}\label{2}
\begin{figure}[tb]
\includegraphics[width=\figwidth]{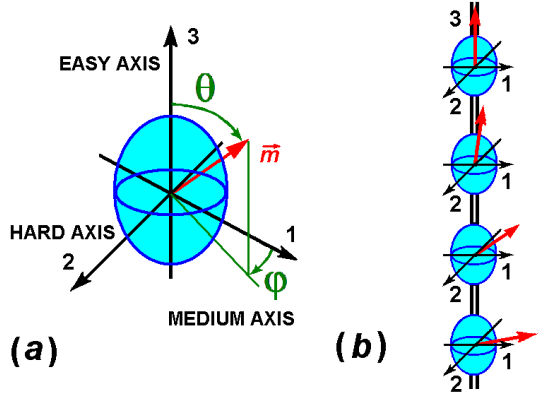}
\caption{The geometry of the problem; $a$) anisotropy axis for a
single magnetic particle; $b$) a schematic representation of a
chain of particles. Each particle is represented by its spin.}
\label{1}
\end{figure}

In order to describe a one-dimensional system of mesoscopic
magnetic particles allowing kink dynamics, we assume that each
particle has an internal magnetic anisotropy, with the chosen axes
to be parallel for all particles in a system. The geometry of the
problem is shown in Fig. \ref{1}. We will also consider isotropic
nearest neighbor interactions. Moreover we assume that any
internal degrees of freedom of the particles can be neglected, and
each particle can be treated as a single magnetic moment (spin).

Note that kinks in a ferromagnet with pure axial symmetry
$\mathrm{C}_{\infty}$ cannot move because the projection of the
total spin on the easy axis $S^{\mathrm{(tot)}}_3$ should change
while the kink is moving. However, the Hamiltonian is invariant
under rotations around the easy-axis and it commutes with
$S^{\mathrm{(tot)}}_ 3$, prohibiting such dynamics. In order to
allow kink dynamics, we can consider the two-fold magnetic
anisotropy. This is in contrast with antiferromagnets, where kinks
can move even for purely uniaxial anisotropy, with an easy axis of
$\mathrm{C}_{\infty}$ symmetry.

\subsection{Model}
The Hamiltonian of a chain-like system of magnetic particles can be
written in the same form as for a discrete ferromagnetic chain
(one-dimensional lattice with atomic spacing $a$) with a spin
operator ${\rm {\bf S}}_n $ located in each lattice site $n$,
\begin{equation}
\label{Sham} \mathcal{H} \;=\;-J \sum\limits_n {\mathbf{S}_{n}
\cdot \mathbf{S}_{n+1}} + \sum\limits_n {[K_1\; S_{1,n}^2 + K_2\;
S_{2,n}^2 ]}.
\end{equation}

Here the first term describes the isotropic exchange interaction
of spins, and the second sum corresponds to the two-fold magnetic
anisotropy. We have chosen $K_2>K_1>0$, so that the orthogonal
axes 3, 1 and 2 are the easy axis, the medium axis and the hard
axis, respectively. We consider each spin operator to be a
classical vector with constant modulus ${\bf S}=S{\bf m}$ and unit
vector ${\bf m}$ (i.e., this vector points in the unit sphere
${\bf m}^2=1$). The dynamics of the variables ${\bf m}_n$ is
governed by the discrete version of the Landau-Lifshitz equation,
\begin{equation}\label{LLE}
\hbar \; S \;\frac{\partial \mathbf{m}_n}{\partial
t}=\left(\mathbf{m}_n \times   \frac{\partial W}{\partial
\mathbf{m}_n}\right),
\end{equation}
where $W \equiv W(\mathbf{m}_n)$ is the energy of the considered
ferromagnetic chain. The dynamics of the variable  $\mathbf{m}_n$,
for a given point $n$ in a lattice, is determined by the Lagrangian,
\begin{equation}
\label{Lagr} \mathcal{L}=-\hbar S \sum\limits_n
{\frac{\mathbf{n}}{1+(\mathbf{n} \cdot \mathbf{m}_n )}\cdot
\left(\mathbf{m}_n \times \frac{\partial {\rm {\bf m}}_n}{\partial
t} \right )}-W,
\end{equation}
where $\mathbf{n}$ is an arbitrary unit vector. For the continuum
approximation, the set of variables $\mathbf{m}_n$ determined on
the lattice sites $n$ should be replaced by a smooth function of
the continuous coordinate $x$: $\mathbf{m}_{n} \to \mathbf{m}(x)$.
In this continuum approximation, the Lagrangian takes the form
\begin{equation} \label{LagrCont} \mathcal{L}=-\int {dx
{\left(\mathbf{A} \cdot \frac{\partial {\rm {\bf m}}}{\partial t}
\right)}}- W, \ \mathbf{A}=\frac{\hbar S}{a} \cdot
\frac{\left(\mathbf{n}\times \mathbf{m}\right ) }{1+(\mathbf{n}
\cdot \mathbf{m} )},
\end{equation}
where $W$ is an energy functional written by expanding over
gradients of $\mathbf{m}$; see, for example, Ref.~\onlinecite{SW}.
The vector $\mathbf{A}$ has the form of a vector-potential of a
Dirac monopole field, i.e., the curl of $\mathbf{A}$ over the
variable $\mathbf{m}$ is proportional to $\mathbf{m}$:
$$\mathrm{curl}_{\mathbf{m}}\mathbf{A}=\hbar S \mathbf{m}/a;$$
see, for example, Ref.~\onlinecite{FRA91}. The vector potential
has a singularity (Dirac string) for
$(\mathbf{m}\cdot\mathbf{n})=-1$, i.e., on a half-line in
$\mathbf{m}-$space. It is important to note that the vector
potential $\mathbf{A}$ is accurate within some gauge
transformations, which includes changing the direction of the
Dirac string, but the Landau-Lifshitz equations containing
$\mathrm{rot}_{\mathbf{m}}\mathbf{A}$ are invariant with respect
to gauge transformations.\cite{FRA91} It is worth mentioning here
that for antiferromagnets within the sigma-model approach, the
dynamical part of the Lagrangian has a standard inertial term
$$L_{\mathrm{dyn, AFM}} \propto \left(\frac{\partial \rm {\bf l}}{\partial
t}\right)^2\ \ ,$$ in contrast to $\mathbf{A} \cdot (\partial {\rm
{\bf m}}/\partial t) $ for ferromagnets.

It is convenient to represent the unit vector field $\mathbf{m}$ by
two independent angular variables $\theta$ and $\varphi$,
\begin{equation} \label{ang}
m_1=\sin\theta \cos \varphi, \ m_2=\sin \theta \sin \varphi, \
m_3=\cos\theta.
\end{equation}
In terms of these variables, the Landau-Lifshitz equation
\eqref{LLE} takes the form
\begin{eqnarray}
\frac{S\hbar }{a}\; \sin \theta\; \frac{\partial \theta }{\partial
t} &=&\frac{
\delta W}{\delta \varphi },  \label{LLEang} \\
\frac{S\hbar }{a}\; \sin \theta\; \frac{\partial \varphi
}{\partial t} &=&-\;\frac{ \delta W}{\delta \theta },
\end{eqnarray}
where $W\equiv W\{\theta ,\varphi \}$ is the energy functional
written it terms of the field variables $ \theta $ and $\varphi $.
The Hamiltonian \eqref{Sham} can be rewritten as
\begin{widetext}
\begin{equation}
W\{\theta, \varphi\}=\int \frac{dx}a\left\{
\frac{Ja^2}{2}\left[\left( \frac{\partial \theta}{\partial
x}\right)+\sin^2 \theta \left(\frac{\partial \varphi}{
\partial x}\right) ^2\right]+ \sin^2 \theta\left[ K_1+\left( K_2-K_1 \right)
\sin^2 \varphi \right] \right\},  \label{energy}
\end{equation}
\end{widetext}
where the non-zero difference $(K_2-K_1)$ determines the
anisotropy in the basal plane.

The Landau-Lifshitz equations have an obvious integral of motion,
the energy $W$. In the continuum approximation, translational
invariance leads to the conservation of the linear momentum of the
magnetization field $P$ (momentum, for short). The expression for
the momentum is determined\cite{Bar-springer,Kosevich+All} by the
dynamical part of the Lagrangian
\begin{equation}\label{P_gen}
P=\int{dx}\left(\mathbf{A} \cdot \frac{\partial \mathbf{m}}{\partial
x}\right).
\end{equation}

As noted above, the vector-potential $\mathbf{A}$ is known up to a
gauge transformation, and the momentum $P$ depends on the gauge
used for $\mathbf{A}$. This is one of the problems for describing
the dynamics of solitons in ferromagnets. This problem does not
exist for antiferromagnets, where the momentum is proportional to
the integral $\int{dx}(\partial \mathbf{l}/\partial x)(\partial
\mathbf{l}/\partial t)$. However, it turns out that the relative
momentum for any \emph{pair of kinks} in a ferromagnet can be
uniquely determined. Thus, the kink momentum is accurate within
the position of the origin in $P-$space; that is, the choice of a
kink assigned with the value of $P=0$.

\subsection{Topological analysis of domain wall structure}
\begin{figure}[tb]
\includegraphics[width=\figwidth]{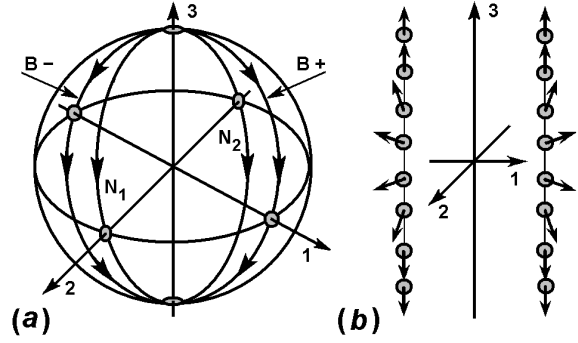}
\caption{(a) Trajectories on the sphere $\mathbf{m}^2=1$ (the
direction of motion is indicated by the arrows) corresponding to
different kinks; the labels 1, 2, and 3 indicate the anisotropy
axes of the magnet. Two of the most favorable kinks with chirality
$\chi=\pm 1$ are shown by the symbols $B+$ and $B-$; unfavorable
kinks with indefinite chirality are labeled by $N_1$ and $N_2$.
The crossings between the trajectories and the axes 1 or 2 are
shown by four gray ovals in (a). The trajectories  $N_1$ and $N_2$
divide the sphere in two domains, associated with kinks of
different chirality, as discussed in the text. (b) Spin
distribution for favorable kinks with chirality  $\chi=\pm 1$.}
\label{f:sfera}
\end{figure}

For a kink in a ferromagnetic chain, the values of the on-site
variables $S_{3,n}$ have opposite values in front and behind of
the kink: $S_{3,n}\rightarrow \pm S$ at $n \to \pm \infty $. In
other words, a kink can be seen as a path connecting the poles of
the sphere ${\bf m}^2=1$, corresponding to the two easy directions
of magnetization space (see Fig.~\ref{f:sfera}). For definiteness,
we assume that $m_{3}= 1$ at $x \rightarrow - \infty $ and $m_{3}=
-1$ at $x \rightarrow +\infty $ (see Fig.~\ref{f:sfera}). For
ferromagnets with non-zero anisotropy in the basal plane,
$\partial W/\partial \varphi \neq 0$, a domain wall can move with
some velocity $v$ smaller than the limit value $v_c$. Within the
continuum approximation, such moving domain walls are described by
a simple traveling-wave one-soliton solution of the
Landau-Lifshitz equation \eqref {LLEang} of the form $\theta
=\theta (\xi )$, $\varphi =\varphi (\xi )$, with $\xi =x-vt$.
However, it is hard to find an analytical solution of a set of two
second-order equations of this type, and we will start with a
qualitative analysis.

Due to symmetry, there are two types of stationary domain walls,
with $\theta = \theta (x)$ and $\varphi = \mathrm{const}$. For these
domain walls, the vector ${\bf m}$ turns either within the easy
plane $(3, 1)$, for the first type of walls, or within the hard
plane $(3, 2)$, for the second one. The trajectories describing
these domain walls are denoted by symbols $B+$, $B-$ and $N_1$,
$N_2$, respectively, on the Fig.~\ref{f:sfera}. Their energies are
$E_{1}$ for
 $B\pm$ and $E_{2}>E_1$ for $N_{1,2}$. Other domain walls having
$v \neq 0$ and energies $E_{1} < E(v) < E_{2}$, are described by
paths, located in between these chosen trajectories on the sphere.

The kink momentum is the total momentum of the magnetization
field, calculated along the corresponding solution of the
Landau-Lifshitz equation.\cite{Bar-springer,Kosevich+All} For
domain walls, it can be written in the form of an integral around
the contour of the sphere Fig.~\ref{f:sfera}, depicting a kink,
$P=\int \mathbf{A}(\mathbf{m})\;d\mathbf{m}$, there $\mathbf{A}$
is the vector-potential of the Dirac monopole field
\eqref{LagrCont}. A difference  of the momentum values for two
different kinks can be described as an integral along a closed
contour. It can be written through a surface integral of the type
of $\int{d\mathbf{S}\; \mathrm{rot}_{\mathbf{m}}\mathbf{A}}$ and
it is equal to $\hbar S/a$, multiplied by the area, on the sphere
$\mathbf{m}^2=1$, inside two trajectories, corresponding to these
pairs of kinks. \cite{galkinaIvPZh00} It is clear that for a
biaxial ferromagnet there are pairs of diametrically opposite
trajectories  (e.g., the trajectories $B+$ and $B-$ in
Fig.~\ref{f:sfera}) corresponding to energetically equivalent but
physically different kinks. For these pairs of kinks, the closed
path borders half of a sphere, with an area of $2\pi $, and the
momentum difference equals $2\pi \hbar S/a$. Hence we can readily
obtain the periodic dependence of the kink energy on its momentum
with the period $P_0$,
\begin{equation}\label{P_0}
P_0=\frac{2\pi \hbar S}{a}.
\end{equation}
For those kinks which are close to the most favorable kink $B+$
($B-$), the value of the momentum $P$ ($\Delta P = P-P_0$) is
small; thus, the parabolic approximation can be used and the
energy can be written as $E=P^2/2M$ (or $E=(\Delta P)^2/2M$),
where $M$ is  the effective mass of the kink. For the model
\eqref{energy}, the effective mass $M$ takes the value of the well
known D\"oring effective mass, obtained as early as the 1930's;
see, e.g., Refs.~\onlinecite{MalozSlon,Kosevich+All}. This
effective mass turns to infinity when $ (K_2-K_1) \rightarrow 0$.
This is another indication that in a pure uniaxial model of a
ferromagnet $(K_2=K_1)$, domain wall motion is impossible. However
there is no contradiction between the finite value of $P \propto
\varphi $ and the condition $v=0$: if $P=Mv$, then the momentum
can be finite when $M\rightarrow \infty $ and $v\rightarrow 0$.

Coherent tunneling assumes the presence of at least two different
states having the same energy; for instance, the two states of
kinks in a biaxial ferromagnet with different values of the
topological charge. A topological
classification\cite{VolovMin77,mermin} of kinks can be done in the
same way for both ferromagnets and antiferromagnets. First, the
difference of values of the magnetization vector ${\bf m}_{x}$ (or
${\bf l}_{x}$, for antiferromagnets) on the right and on the left
sides of the kink determine the topological charge $\pi_0$ of the
kink. Changing this topological charge requires overcoming the
potential barrier, proportional to the system size (formally,
infinite barrier) that cannot be realized by tunneling. Second, a
topological charge of the type $\pi_1$ is determined by mapping
the coordinate space of the spin chain (the line $-\infty < x <
\infty $) onto the circle $\{m_1^2+m_3^2=1,m_2=0\}$, situated in
the easy-plane of the ferromagnet. The appearance of two
topological charges of different levels can be formally described
using the relative homotopy group, as discussed in
Ref.~\onlinecite{mermin}.

The $\pi_1$ charge is described by the integral
$$\chi=\frac{1}{\pi}\int^{\infty}_{-\infty} \left({\bf e}_2\left({\bf m}
\times \frac{\partial{\bf m}}{\partial x}\right)\right)dx.$$  In
other words, the \emph{chirality} $\chi=\pm1$ determines the sense
of rotation (clockwise or counterclockwise) of ${\bf m}$ along the
chain. This standard definition of the chirality used in Refs.~
\onlinecite{BraunLoss96,TakagiTatara96,ShibTakagi00} is only valid
for kinks located on the unit sphere near the most energetically
favorable configurations $B+$ and $B-$, as shown in
Fig.~\ref{f:sfera}; that is, for kinks having small velocities
$v$. For this case, the effective mass approximation is valid, and
these two kinks are well separated. According to the topological
analysis, these kinks have different values of the chirality $\chi
= \pm 1$; and within a self-consistent Hamiltonian approach, they
correspond to different values of the momentum: $P=0$ and $P=P_0$.
However, for kinks moving with a non-small velocity and with
arbitrary values of the momentum, the above definition of the
chirality should be modified.

For treating the whole order parameter space (i.e., the sphere
${\bf m}^2=1$), kinks with $\chi=\pm1$ can be transferred to each
other through energetically unfavorable kinks of type $N_1$ or
$N_2$, schematically shown in Fig.~\ref{f:sfera}. Here the barrier
is finite (and equal to $E_{2} - E_{1}$; which is large when $K_2
\gg K_1$) and the process of kink chirality $\chi$ tunneling is
possible. The concept of chirality, as a discrete number $\chi =
\pm 1$, is naturally connected with the presence of a discrete
degeneracy in the dependence of the kink energy on its momentum.
The discrete parameter $\chi = \pm 1$ determines one of à two
different, but energetically equivalent, kink states existing in a
biaxial ferromagnet. The values $\chi=1$ and $\chi=-1$ can be
naturally attributed to kink states with trajectories in two
equivalent semi-spheres, $m_1>0$ and $m_1<0$, respectively. The
chirality value is not determined for the unfavorable static kinks
only (Neel walls) for which the trajectories $N_1$ and $N_2$ pass
through the hard axis. In this sense, chirality tunneling can be
seen as a tunneling effect in momentum space with a non-small (of
the order $P_0 =2\pi \hbar S/a$) change of the kink momentum.

\subsection{Moving domain wall structure}
To confirm the general features for moving domain walls discussed
above, we will discuss an exact solution of \eqref {LLEang} known
for the model of biaxial ferromagnets with the energy
\eqref{energy}. It is easy to find the structure of domain walls
with zero velocity. There are two types of domain walls having
thickness $x_{1,2} = a\sqrt{J/2K_{1,2}}$ and energies $E_{1,2} = 2
S \sqrt{2J K_{1,2}} $ with $E_{1} < E_{2}$. These are the
one-dimensional analogs of the usual Bloch and Neel domain walls;
see Refs.~\onlinecite{MalozSlon,Kosevich+All}.

The structure of a domain wall moving with a non-small velocity
within the model \eqref{energy} was obtained by Walker at the end of
the 1950s, see
Refs.~\onlinecite{MalozSlon,Bar-springer,Kosevich+All}. For this
solution, the function $\theta =\theta (\xi )$, $\xi =x-vt$, and the
value of $\varphi = \varphi _0= \mathrm{const}$ is independent on
$\xi$. The value of $\varphi _0$ is determined by the domain wall
velocity $v$ as follows
\begin{equation}
\frac{v\hbar }{a\sqrt{JK}}=q\frac{\varepsilon \sin \varphi \cos
\varphi }{ \sqrt{1+\varepsilon \sin ^2\varphi }},   \label{fi(v)}
\end{equation}
here and below we use the notation $\varepsilon=(K_2-K_1)/K_1$  to
shorted the expressions. The relation \eqref{fi(v)} governs, in
particular, the maximal possible value of a domain wall velocity,
the so-called Walker velocity $v_W$
\begin{equation}
v_W=(a/\hbar )\sqrt{JK}(\sqrt{1+\varepsilon }-1).  \label{v w}
\end{equation}
It is interesting to note that $v_W$ is smaller than the minimal
phase velocity of spin waves, $$v_{\mathrm{ph}}=(a/\hbar
)\sqrt{JK}(\sqrt{1+\varepsilon } +1).$$ The values of $v_W$ and
$v_{\mathrm{ph}}$ coincide only in the limit $ \varepsilon
\rightarrow \infty ,$ when the Landau-Lifshitz equation (\ref
{LLEang}) can be mapped onto the sine-Gordon equation. However, if
$\varepsilon $ is small, then $$v_W\simeq \varepsilon
v_{\mathrm{ph}}/4 \ll v_{\mathrm{ph}}.$$ The value of $v_W$
vanishes when $\varepsilon \rightarrow 0$; that is, as was noted
above, the domain wall cannot move at all for pure uniaxial
ferromagnets.

The Walker's solution can be presented in the explicit analytical
form
\begin{equation}
\cos \theta =q \cdot \tanh \left[\frac{\xi}{x_0 (v)}\right], \
\sin \theta =\sigma\left\{\cosh \left[\frac{\xi}{x_0
(v)}\right]\right\}^{-1},
 \label{walker}
\end{equation}
where $q =\pm 1$,  $\sigma =\pm 1$,
\begin{equation}
x_0(v)=a\sqrt{\frac{J}{2K_1(1+\varepsilon \sin^2 \varphi)}}
 \label{xotv}
\end{equation}
and $\varphi = \varphi (v)$ is determined by Eq.~\eqref{fi(v)}.
The two topological charges introduced above naturally appear here
as the quantities $q$ and $\sigma $;  $q =\pm 1$ determines the
$\pi _0-$topological charge of the kink, and $\sigma =\pm 1$
governs the spin direction in the kink center and it is naturally
connected with the $\pi _1-$topological charge, the chirality
$\chi $.

A straightforward calculation of the kink energy leads to the
formula
\begin{equation}
E=E_1\sqrt{1+\varepsilon \sin ^2\varphi },\ E_1=2S^2\sqrt{2JK_1},
\label{e ot fi}
\end{equation}
where $\varphi =\varphi (v)$ is determined by Eq. (\ref{fi(v)}).
The $E(v)$ dependence consists of two branches. In
three-dimensional ferromagnets, where two-dimensional plane domain
walls are present, the upper branch of the domain walls is
unstable. However this instability is developed via non-uniform
perturbations in the domain wall plane, and such fluctuations do
not exist for domain walls in one-dimensional magnets. Below we
will show that for the more natural domain wall energy
representation (namely, as a function of its momentum) the $E(P)$
function is single-valued.

For the Walker solution, the value of $\varphi$ is
$\xi-$independent,\cite{MalozSlon,Bar-springer} and the paths on
the sphere present at Fig.~\ref{f:sfera} are the halves of the big
circles passing through the poles of the sphere. Therefore, the
domain wall momentum can be written as
\begin{equation}\label{WalImp}
   P=P_0\frac{\varphi}{\pi},
\end{equation}
and the explicit form of the dependence of the domain wall energy on
its momentum can be rewritten as
\begin{equation}
E=E_1\sqrt{1+\frac{2T_0}{E_0} \sin ^2 \frac{\pi P}{P_0} }, \ \
 \frac{2T_0}{E_1} = \frac{K_2-K_1}{K_1}
\label{EotP}
\end{equation}
where $$E_1=2S^2\sqrt{2JK_1}$$ is the minimum energy of the domain
wall introduced above. It is worth noting that the equation
\eqref{EotP} reproduces the dependence $v(\varphi)$ \eqref{fi(v)}
within the Hamilton dynamics, $$\frac{dX}{dt} \equiv  v =
\frac{\partial H(X,P)}{\partial P}.$$ This leads to unusual
dynamical features, like the oscillatory motion of a domain wall
as a response to a dc driving force (e.g., dc magnetic field
parallel to the easy axis). These peculiarities are well-known for
the exact Walker solution and have been experimentally established
for moving domain walls in magnetic bubble materials; see, e.g.
Ref.~\onlinecite{MalozSlon}. Here we were able to write down the
explicit form of the function $E(P)$, but the periodic dependence
$E(P)$ with the same value of $P_0$ is present for any continuum
model of a ferromagnet with biaxial anisotropy. Such periodic
dependence $E(P)$  is also valid for biaxial discrete models; for
details, see Ref.~\onlinecite{galkinaIv07}.

\subsection{Kink coordinate and lattice pinning}
As we will show, the quantum properties of kinks can be described
within a semiclassical analysis of the Hamilton dynamics of
collective variables: the kink coordinate $X$ and conjugated
momentum $P$. This dynamics is determined by the characteristic
Hamilton function $H(P,X)$. In a continuous approximation, the
definition of kink coordinate is obvious. However, the Hamilton
relation $$\frac{dP}{dt}\;=\;-\;\frac{\partial H(P,X)}{\partial
X}$$ shows that the kink momentum is conserved for any model with
the Hamiltonian independent on $X$. Therefore any processes of
tunneling in momentum space (in particular, the tunneling of the
domain wall chirality) can only occur  if the Hamilton function
$H$ depends on the domain wall position $X$. For our model, such
dependence can only be caused by a lattice pinning of the kink.
Thus, for a consistent description of quantum tunneling, lattice
pinning must be considered.

A first step in this direction is to define the domain wall
coordinate $X$ treated as a collective variable and conjugated to
the kink momentum $P$. The kink coordinate $X$ in the discrete
model can be naturally determined through the spin operators, $$X
= \frac{a}{2S}\sum\limits_n {[S_{3,n} -S_{3,n}^{(0)} ]}$$, where
$S_{3,n}^{(0)} $ corresponds to a certain ``reference''
kink,\cite{IvanovMik04} which coordinate is chosen as $X=0$. The
total spin projection onto the axis 3 is conserved,
$S_3^{\mathrm{tot}}=\sum\limits_n S_3= \mathrm{const}$; thus, for
uniaxial ferromagnets with $K_{2}=K_{1}$, $dX/dt=0$ and a kink
dynamics is impossible. Another consequence appears when taking
into account the Hamilton relation $dX/dt =
\partial H(P,X)/\partial P$.  For the purely uniaxial case, $H(P,X)$
does not depend on $P$. All of these general considerations here are
characteristic of  the exact Walker solution.

A spin configuration corresponding to a kink with a specified
value of the coordinate $X$ can be obtained by minimizing the
Hamiltonian with respect to the variables $\theta_n$ and
$\varphi_n$, for a fixed value of the total spin
$S^{\mathrm{(tot)}} _3$. To do this, we use a procedure proposed
and numerically realized for the analysis of different dynamical
solitons,\cite{Iv+PRB06} which are described by a conditional
minimum of a discrete spin Hamiltonian. Using this method one can
easily determine the structure of the kink and obtain the
dependence of the kink energy on its coordinate $X$; this for
finite spin chains described by any classical spin Hamiltonian.

Now, we  consider model \eqref{Sham} with a purely single-ion
anisotropy, for a finite chain of size $N_{c}$, with boundary
conditions $\cos \theta _n = 1$ and $\cos \theta _n = -1$ at
different ends of the chain. The size of the chain $N_{c}$ is
chosen to be much larger than the width of the kink. In
particular, for a reasonable anisotropy $K > 0.2J$, the kinks
occur as well-localized excitations. As a result, the kink energy
is independent of $N_{c}$ for $N_c \geq 30$. For extremely high
values of the anisotropy, $K
> K_c$, with $K_c = 0.667J$,  the domain wall becomes
purely collinear,\cite{collDW} with all spins up or down, $S_3 = \pm
S $. For such collinear states, the continuum description of the
domain wall dynamics and its topological analysis are obviously
incorrect, and we should restrict our consideration to moderate
values of the anisotropy: $0.2J < K \leq 0.65J < K_c$.

\begin{figure}[tb]
\includegraphics[width=\figwidth]{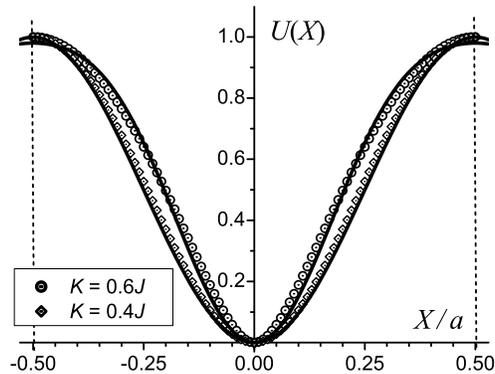}
\caption{Shape of the domain wall pinning potential $U(X)$,
normalized by its maximal value $U(a/2)$, for two values of the
anisotropy constant. Symbols denote the numerical data. The fit
for the model dependence \eqref{u(x)2} is shown by the full line.}
\label{f:UotX}
\end{figure}

Considerable influence of the lattice pinning appear when $K \geq
0.25J$. A more favorable position of the kink is between neighboring
spins. Thus, the values $S_{3,n}$ at two neighboring spins are equal
in magnitude and opposite in sign: $S_3 = \pm S(0) < 1$. Choosing
the value of $X=0$ for one of such states, we can determine the
pinning potential having equivalent minima at the points $X = an$,
where $n$ is an integer. The states with kink on a lattice site with
$X = a(2n+1)/2$ correspond to maxima of the pinning potential, as
shown in Fig.~\ref{f:UotX}. In general, it can be concluded that,
for moderate values of the anisotropy $K \leq K_c$, the pinning
potential is not large compared to the ``static'' energy of the
kink. The dependence $U(X)$ is fairly well described by the simple
harmonic relation
\begin{equation} \label{u(x)1}
U(X)= U_0 \cdot \sin ^2\left(\frac{\pi X}{a}\right)\ ,
\end{equation}

\begin{figure}[tb]
\includegraphics[width=\figwidth]{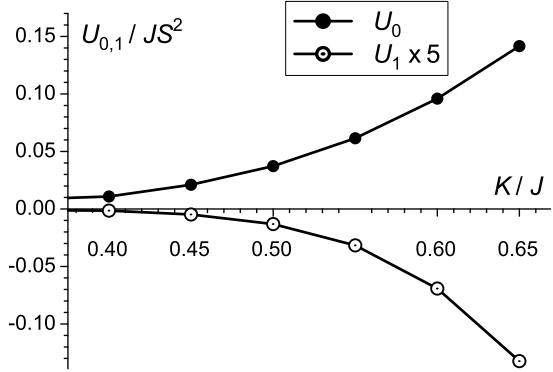}
\caption{Coefficients $U_0$ and $U_1$ (in units of $JS^2$) in Eq.
\eqref{u(x)2}  for some values of anisotropy constant $K_1$.}
\label{f:U12}
\end{figure}

On the other hand, higher Fourier components are also present in
the dependence $U(X)$, especially for higher anisotropy. For
example, for a more general form
\begin{equation}
\label{u(x)2}U(x)=U_0\sin ^2\left(\frac{\pi X}{a}\right)+U_1\sin
^2\left(\frac{2\pi X}{a}\right),
\end{equation}
the contribution of $U_1$ can be noticeable. The values of $U_0$
and $U_1$ as functions of the reduced anisotropy $K/J$ are
presented in Fig.~\ref{f:U12}. In general, for higher values of
the anisotropy $K$ one can see a broadening of the curve $U(X)$
near its maximum and, correspondingly, a narrowing of the curve
near the minima. Note that when $K>K_c$, the function $U(X)$ has a
cusp at $X=0$.

To conclude this section, we derive  the periodic dependence of
the domain wall Hamiltonian on both collective variables $X$ and
$P$. Namely, to describe the quantum dynamics, we can use the
Hamiltonian $H(P,X)=T(P)+U(X)$, where both functions $T(P)$ and
$U(X)$ are periodic: $U(X)=U(X+a)$ and $T(P)=T(P+P_0)$.

\section{QUANTUM TUNNELING EFFECTS IN KINK DYNAMICS}\label{3}
To describe the quantum dynamics of domain walls,  we can proceed
with the Hamiltonian $H(P,X)=T(P)+U(X)$, where the ``kinetic
energy'' $T(P)$ is described by  Eq.~\eqref{EotP} and the
``potential energy'' is caused by a periodic lattice pinning
potential $U(X)=U(X+a)$; see Eqs.~\eqref{u(x)1} or \eqref{u(x)2}.
The most crucial point is the presence of the double periodicity of
$H(P,X)$, with respect to both $X$ and  $P$. For simplicity, below
we use the simplest model holding this property
\begin{equation}
\label{HamPX} H=T_0 \sin ^2\left(\frac{\pi P}{P_0 }\right)+U_0 \sin
^2\left(\frac{\pi X}{a}\right),
\end{equation}
where only the lowest harmonics are considered. Here the energy of
the kink is taken from its minimum classical value, i.e., the energy
of a ``unmoving'' kink with $P=0$ or $P=P_0$, situated at the
minimum of a pinning potential $X=0$.

For different ferromagnetic chains, any ratio of parameters $T_0$
and $U_0$ is in principle possible. For example, the value of
$T_0$ is proportional to the difference $(K_2-K_1)$ and it
vanishes in the purely uniaxial case for any value of $K_2=K_1\neq
0$, while the amplitude of the pinning potential $U_0$ is almost
zero for $K_1 < 0.15 J$. It is natural to start with the
Bohr-Sommerfeld quantization for the domain wall motion, which is
based on the analysis of the classical dynamics. This can be done
in the same way as for the transverse-field Ising model (see Ref.
\onlinecite{IvanovMik04}) and we do not discuss its details here.
The most important feature of a Hamiltonian of the type
\eqref{HamPX} is  the presence of a lower and upper bound for the
energy. Hence, two types of finite motion appear. A first type
corresponds to oscillations of the domain wall with either $P \ll
P_0$ or $(P-P_0) \ll P_0$ near the minimum of the potential, with
the energy near the minimum of the Hamiltonian, $E \ll T_0$,
$U_0$. A second type of finite motion corresponds to oscillations
near the potential maximum, with the momentum near the values of
$P=\pm P_0/2$, the values of energy are $E \leq T_0 + U_0$.

For intermediate values of the energy, the motion is infinite. For
a small pinning potential, this motion is standard, with infinite
grow of the kink coordinate. The case $U_0 > T_0$ is less
standard; it corresponds to an infinite-growing momentum with
finite oscillations of the coordinate near certain positions,
which do not coincide with extrema of the pinning potential. The
late case is nothing but Bloch oscillations in the pinning
potential $U(X)$. An exception is the chosen point $T_0 =U_0$, for
which the classical motion is finite for all values of the energy.

Both types of infinite motion describe the classical over-barrier
dynamics of the domain walls. Using a quantum-mechanical language,
such states of the nearly-free particle can be well described by
perturbation theory over $U_0$ or $T_0$, for the cases $U_0 <T_0$ or
$U_0>T_0$, respectively. This analysis will be done in the next
subsection \ref{PT}.

Within the Bohr-Sommerfeld quantization condition, all the states
corresponding to finite motion (oscillations) of the domain wall
near any extrema of the Hamiltonian produce a discrete spectrum of
energy levels $E_n$ with a level separation of the order of
$\sqrt{U_0T_0}/S$. Both $U_0$ and $T_0$ are proportional to $S^2$,
therefore for the semiclassical situation of high spins $S \gg 1$,
the separation of values of $E_n \propto S$ and it can be smaller
than $U_0$ or $T_0$. These states, with energy $E_n$, are well
localized. For them, the probability of tunneling is small, and
for its estimate the semiclassical approximation is adequate. Such
analysis will be done in subsection \ref{inst}.

\subsection{Perturbative analysis}\label{PT}
For extremely large or small values of $U_0/T_0$ (namely, for
$U_0/T_0<1/S^2$ or $T_0/U_0<1/S^2$) the value of  $E_{n+1} - E_{n}
\sim \hbar \omega _{n} $ can be of the order of $\min[T_0, U_0]$.
In these cases, all the states are delocalized, and our
perturbation theory gives the full description of the domain wall
spectrum. Out of these strong inequalities, a perturbative
analysis can only be applied to domain wall states with
intermediate values of the energy, which correspond to the
classical infinite motion discussed above.

Let us now start with the case $U_0 \ll T_0 $, using perturbation
theory with respect to $U(X)$. In this case, in zeroth
approximation, $P=\mathrm{const}$. To proceed further, let us
assume the chain to have a large but finite size $L=Na$, $ N \gg
1$. Then periodic boundary conditions give the usual
quasi-continuous spectrum of the momentum $P=P_n=(2\pi \hbar
/a)(n/N)$,  where $n$ is an integer, $N/2<n<N/2$, or
$-P_B/2<P<P_B/2$, where $P_B=(2\pi \hbar /a)$ is the size of the
usual (crystalline) Brillouin zone. The vector of states
$|P\rangle$ corresponds to a fixed value of momentum and, hence,
the fixed value of chirality and the uncertainty value of the kink
coordinate. The quantum spectrum of the problem repeats the
dependence of the Hamiltonian (\ref{HamPX}) on $P$.

We now consider the term $U(X)$  as a perturbation. Its role will
generally be the same as for the lattice potential $U(X)$  in the
standard weak-binding approximation in solid state physics. At
zeroth order approximation in the coordinate space, the
eigen-functions are of the form  $\psi ^{(0)}=\exp (iPX)$, with
the energy $E^{(0)}(P)=T(P)$. The influence of the potential
$U(X)$ with the period $a$ leads to the formation of Bloch states
which are a superposition of the states $\psi ^{(n)}=\exp
(iPX+inP_B )$, and the momentum transforms to quasi-momentum. In
the weak-binding approximation, the spectrum can be obtained by a
superposition of unperturbed dispersion curves $E^{(0)}=T(P)$,
with argument shifting by $nP_B $, where $n$ is an integer number.
This spectrum is periodic with the period equal to the size of the
first Brillouin zone $P_B$. The influence of the perturbation is
maximal if the values of the functions $(P+nP_B )$ and
$(P+n^{\prime}P_B )$, with the different $n \neq n^{\prime}$,
coincide for some value of $P$.

In contrast with Bloch electrons with a parabolic dispersion law
$E^{(0,el)}=P^2/2M$, for kinks in ferromagnets, the unperturbed
dispersion law is already described by a periodic function. Hence,
for the resulting dispersion law $E(P)$, the periodic dependence
(with the period matched with both characteristic values $P_0
=2\pi S\hbar /a$ and $P_B =2\pi \hbar /a$) should appear. It is
also important that kinks for the states $|P\rangle$ and
$|P+P_0\rangle$ have the same energy and velocity, but differs by
the sign of the chirality $\chi=\pm 1$. In pure classical
language, these states are described by different magnetization
distributions, their images corresponding to diametrically
opposite paths on the sphere in Fig.~\ref{f:sfera}.

\begin{figure}[tb]
\includegraphics[width=\figwidth]{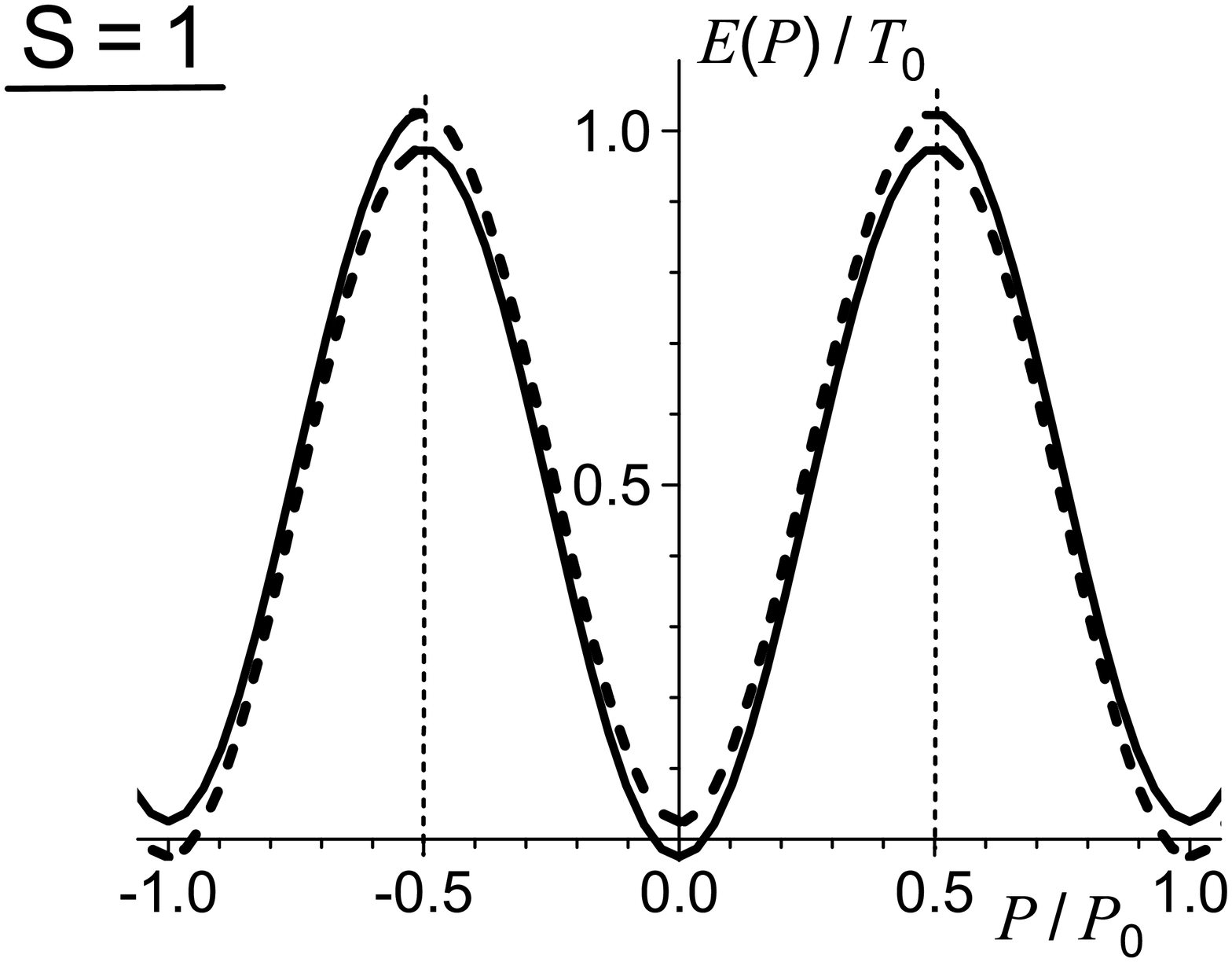}
\caption{ \label{S1} Dispersion relation of a kink in a
ferromagnet with spin $S=1$ subject to a weak pinning potential.
The solid line and the dashed line correspond to the states
$(|+\rangle - |-\rangle)$ and $(|+\rangle + |-\rangle)$ ,
antisymmetric and symmetric over chirality, respectively.  Here,
and in the Figs.~\ref{S2}--\ref{S3to2}, the vertical dotted lines
show the boundaries of the usual Brillouin zone, $-P_B/2 < P\leq
P_B/2$.}
\end{figure}

A simple analysis shows the fundamental difference between the
character of the spectrum for integer and  half-integer  values of
the atomic spin $S$. For minimal integer $S=1$,  the periods $P_0
$ and $P_B $ coincide: see Fig.~\ref{S1}. The accounting of the
potential $U(X)$ of the form (\ref{HamPX}) leads to the
overlapping of functions $E^{(0)}(P)$ and $E^{(0)}(P+P_0)$.
Taking, for definiteness, a $P$ situated in the first Brillouin
zone,  $- P_0/2<P\leq P_0/2$, we can say that these unperturbed
states have a different chirality $\chi=+1$ and $\chi=-1$. The
action of the potential leads to their hybridization and formation
of the states $|P\pm\rangle=(|P\rangle\pm
|P+P_0\rangle)/\sqrt{2}$, having energies
$$E^{(\pm)}(P)=E^{(0)}(P)\pm \langle P|U(X) |P+P_0\rangle,$$ where
$$E^{(0)}(P) = T(P)+\langle U(X) \rangle,$$
$$\langle P|U(X)
|P+P_0\rangle)=U_0/4$$ and $$\langle U(X) \rangle \equiv \langle
P|U(X) |P\rangle)=U_0/2$$ are the off-diagonal matrix element and
the mean value of the potential $U(X)$ for (\ref{HamPX}),
respectively.

\begin{figure}[tb]
\includegraphics[width=\figwidth]{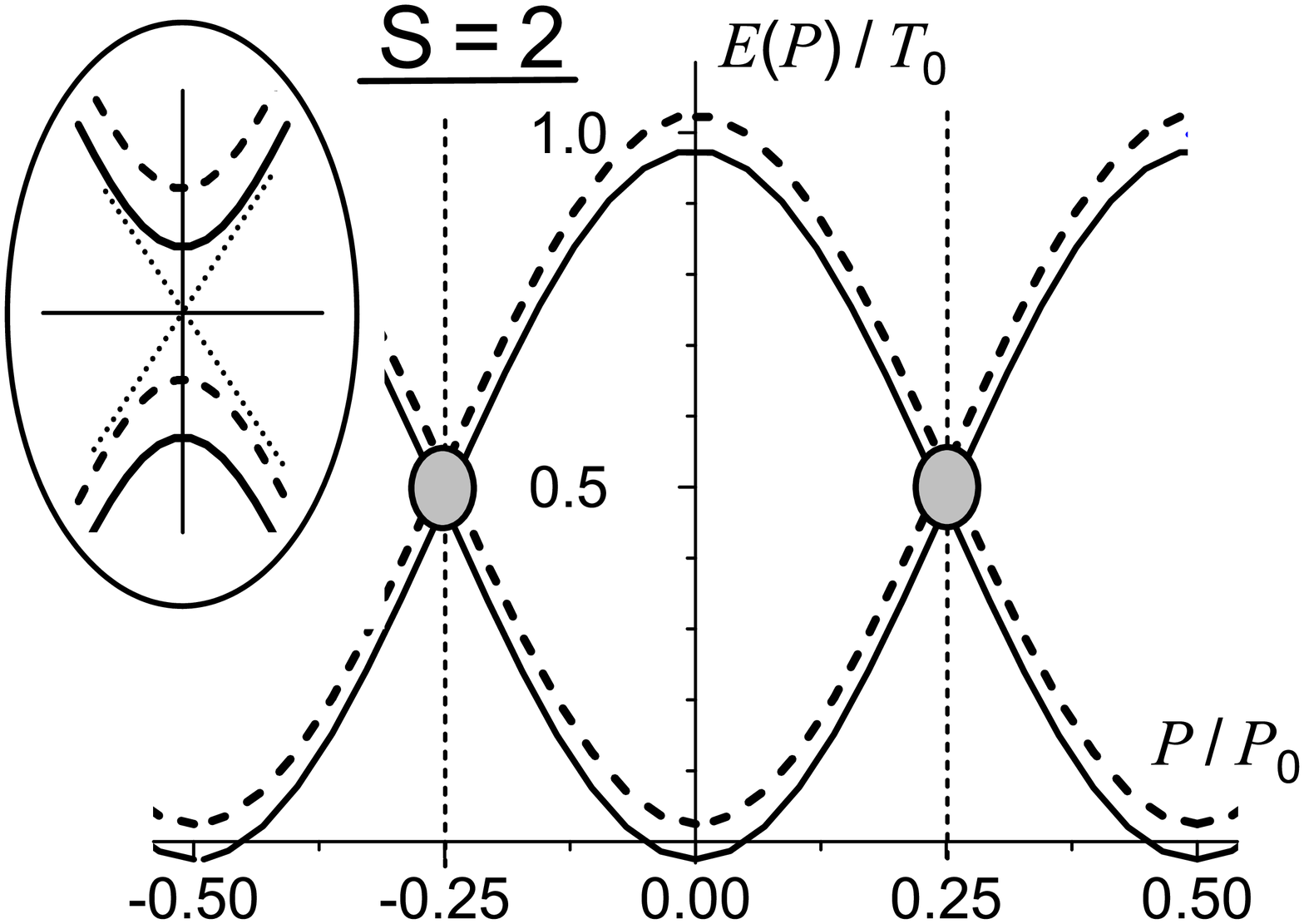}
\caption{ \label{S2} Same as for Fig.~\ref{S1} for spin value
$S=2$. The region near the crossing points, schematically shown by
shadowed ellipses on the main figure, are magnified in the insert
upward left.}
\end{figure}

The same expression can be found for any integer spins $S=k>1$, i.e.
the Fourier component of the potential with $P=kP_B=P_0$ leads to a
full hybridization of the chirality for any $P$. As a result, states
of type $|+\rangle \pm |-\rangle$ appear, where $|\pm\rangle$
correspond to chirality values $\chi=\pm 1$. Such states are the
quantum superposition of the kinks describing diametrically opposite
trajectories on the sphere and energy $E^{(\pm)}(P)=E^{(0)}(P)\pm
U_0/4$, where $U_0$ is the corresponding matrix element of the
potential $U(X)$.
For integer spins $S>1$, the value $P_0 =S\cdot P_B $ is a common
period. Neglecting the chirality tunneling, one can find the $S$
usual energy bands (doubly degenerated over the chirality values)
with the size of $P_B $. The chirality tunneling splits any of
them into two subbands, corresponding to states $(|+\rangle \pm
|-\rangle)$ with $E^{(\pm)}(P)$, and the total number of bands
equals to $2S$, as shown in Fig.~\ref{S2}.

\begin{figure}[tb]
\includegraphics[width=\figwidth]{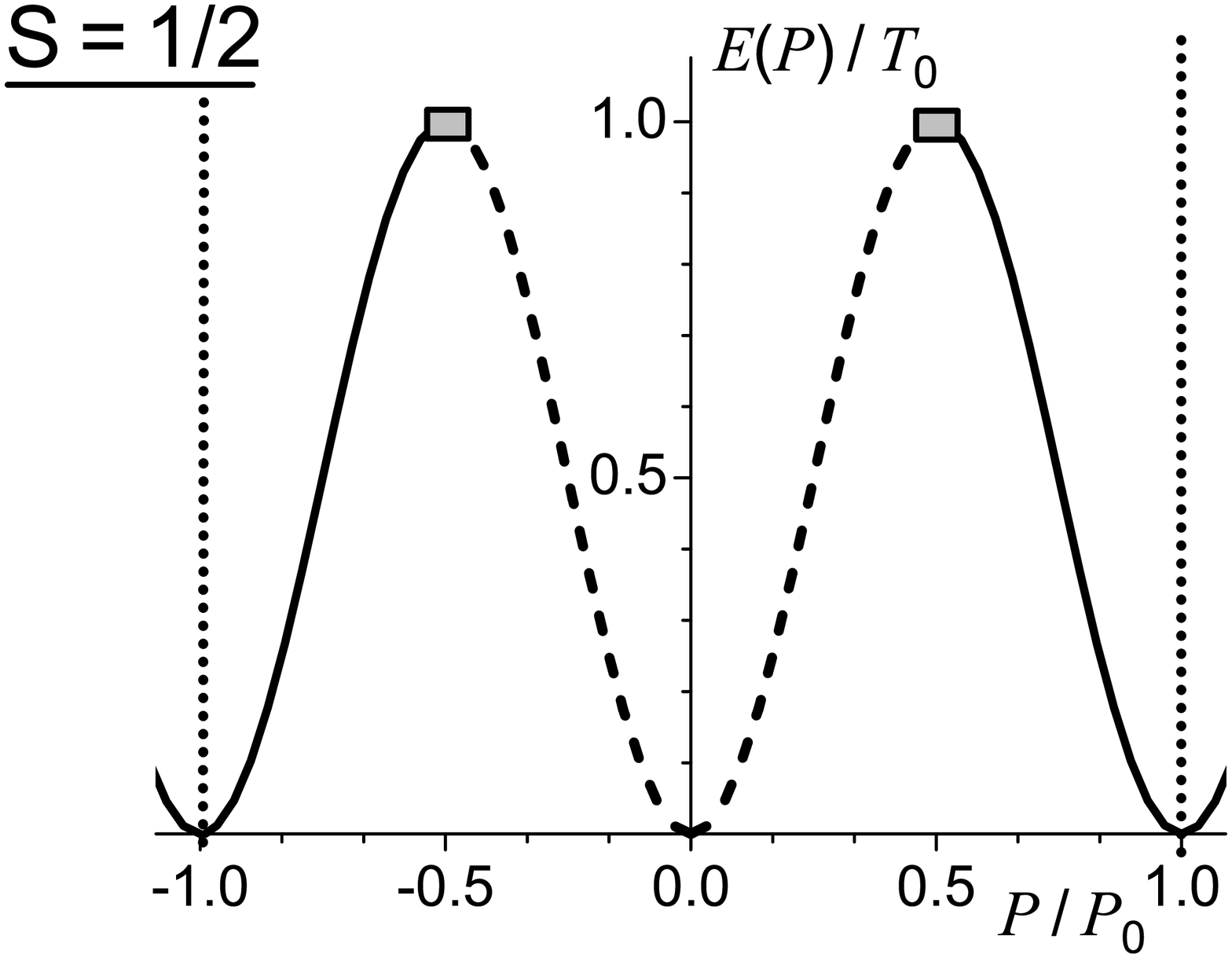}
\caption{\label{S1to2} Dispersion relation of a kink in a
ferromagnet with spin  $S=1/2$,  subject to a weak pinning
potential. The solid line and dashed line correspond to the states
with chirality $\chi =-1$ and $\chi =1$, respectively. Rectangles
denote the points where the value of the chirality is not
determined. }
\end{figure}

\begin{figure}[tb]
\includegraphics[width=\figwidth]{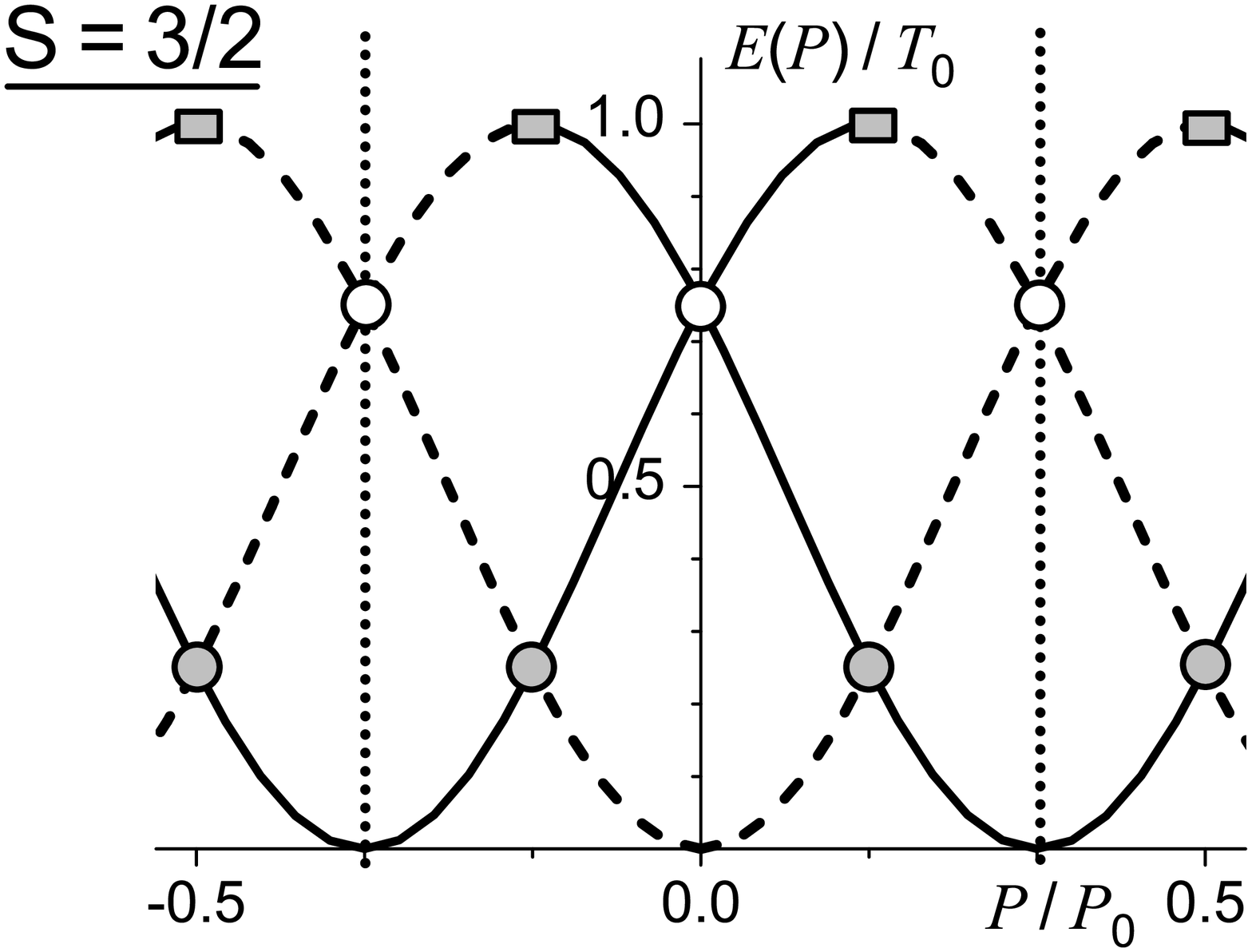}
\caption{\label{S3to2} Same as in Fig.~\ref{S1to2} for the spin
value ${S}=3/2$.  The gray and the light circles schematically show
the areas in the vicinity of the crossing of unperturbed spectra
with and without chirality hybridization, respectively. }
\end{figure}

For half-integer spins $S=k+1/2$, where $k$ is an integer number,
the situation is completely different. It is easy to show that
none of the Fourier components of the potential with
$nP_B=2nP_0/(2k+1)$ leads to such an overlapping of the
non-perturbed spectrum at any $P$, which take place for integer
$S$. For $S=1/2$ any crossing is absent,  as shown in
Fig.~\ref{S1to2} [a crossings of non-perturbed spectra for spin
$S=1/2$ reported in Ref.~\onlinecite{BraunLoss96} is an artifact
of the parabolic approximation for $E^{(0)}(P)$]. For higher
half-integer spins $S>1/2$ the only crossings at some fixed values
occur when $P=P_n$. Such crossings can appear for branches
$E^{(0)}(P+nP_B)$, corresponding to the kinks with the same or
different chirality, as shown in Fig.~\ref{S3to2}. For this last
case, the effects of chirality hybridization can be present very
near the crossing points, $|T(P)-T(P_n)| \ll U_0$, as shown in
Fig. \ref{S3to2}.

Let us consider the opposite limiting case  $U_0 \gg T_0$, when
the kinetic energy $T(P)$ plays the role of a small perturbation
for the unperturbed Hamiltonian $H_0=U(X)$. To zeroth order
approximation with respect to $T_0$, the solution is now
$X=\mathrm{const}$. To construct this perturbation theory, the
momentum representation should be used. In this case, taking into
account the identity of the states with $P$ and $P+2P_0$, we
should apply the condition $\exp[i(P-P_0)X] = \exp[i(P+P_0)X]$,
that leads to the discreteness of the values of the kink
coordinate, $X=X_k=ak/2S$. The difference between this boundary
condition and the condition which was used for the case $U_0 \ll
T_0$ above, leads to essentially different results.

For the case of strong pinning potential $U_0 \gg T_0$, unperturbed
states are described by the wave function in the momentum
representation $\psi ^{(0)}=\exp(iX_kP)$, with definite coordinate
$X=X_k$ and indefinite value of the momentum, which also means an
indefinite value of the chirality. Let us now consider $T(P)$ as a
small perturbation. It is easy to see that $n$th Fourier component
of $T(P)$, with $\exp(2i\pi n P/P_0)$, leads to a nonzero matrix
element of the quantum transition when changing the kink coordinate,
$X_k\rightarrow X_k+\delta X_n$, $\delta X_n=na/S$ (in the simplest
case with one harmonic only, Eq.~\eqref{HamPX}, the transition with
$n>1$ requires accounting the $n$th order in perturbation theory).
If $U(X_k)=U(X_k+\delta X_n)$,  i.e. when the value of $\delta X_n$
is multiple to the chain period $a$, the ``resonant'' transition
should be observed.

Summarizing, kinks in ferromagnetic chains with either integer or
half-integer spins behave differently. The tunneling from a cell
to the neighboring one is possible for an integer spin, while for
the half-integer spin only the transitions with changing $X$ by
$2a$ are permitted, i.e. ``jumps'' across one cell. This feature
was mentioned in Ref.~\onlinecite{BraunLoss96} using a different
reasoning.

\subsection{Semiclassical dynamics}\label{inst}
For large values of the spin, $S \gg 1$, a semiclassical approach
provides a much better approximation than the perturbative approach
used above. For electronic states in a large lattice potential, the
semiclassical tunneling can be formally described by the
tight-binding approximation, which leads to the formation of a band
spectrum of the kink with narrow allowed bands.

Considering possible tunneling effects for a kink, one immediately
encounters the question of under-barrier transition in both
coordinate space and momentum space. For domain walls with the
minimal energy, such transitions include the tunneling between
states corresponding to  a two-dimensional set of points in phase
space of a system $(P,X)$, such as  $X\simeq 0, \pm a, \dots $ and
$P\simeq 0, \pm P_0, \dots $. Within the semiclassical
approximation, these transitions can be investigated in the
framework of the instanton approach; see
Refs.~\onlinecite{InstAlphabet,QTM95}. This approach is a version of
the Feynman path integral method suited to the description of the
underbarrier transitions. It involves using the Euclidean
space-time, that is transforming to the imaginary time, $t \to i\tau
$ (so called Wick rotation). Within this approach, the amplitude of
the underbarrier transition from a given quantum state $|i\rangle$
to another one $|f\rangle$ is determined by the path integral $\int
{D\!X\cdot \exp [-A_E[X] /\hbar ]} $, where $D\!X$ denotes
integration over all possible paths that satisfy the specified
boundary conditions. Here the Euclidean action $A_E[X]$ is described
in the form  $A_E =\int {L_E d\tau }$, and $L_E=P(dX/d\tau)-H(P,X)$
is obtained by the application of a Wick rotation to the usual
mechanical Lagrangian. The instanton solution determines the
trajectory for which the tunneling amplitude is maximal, i.e.; the
instanton trajectory minimizing $A_E$ with respect to $X(\tau )$ and
$P(\tau )$, with the conditions $|i\rangle$ at $\tau \to - \infty$
and $|f\rangle$ at $\tau \to +\infty$. The minimum of the Euclidean
action is realized on the separatrix solution of the corresponding
Euler-Lagrange problem for the Euclidean action functional $A_E$ or,
equivalently, on the solution of the Hamilton equation with the
substitution $t\to i\tau $. The tunneling splitting of the levels
$\Delta$ is determined by the formula $\Delta =\eta \hbar \omega_0
\sqrt{A_E/\hbar} \exp(-A_E^{0}/\hbar) $, where $\omega_0$ is the
characteristic frequency, $\eta \sim 1$.

Let us apply this approach to the mechanical problem of the kink
dynamics described by the Hamiltonian \eqref{HamPX}. It is easy to
see that the Wick rotation $t\to i\tau $, simultaneously with the
simple substitution  $X \to i\Xi $, reduces the instanton problem to
the Hamilton problem for \emph{real} canonical variables $\Xi$ and
$P$ and with the \emph{real} Hamilton function $H_E $,
\begin{equation}
\label{Xi} {H}_E =T_0 \cdot \sin ^2(\pi P/P_0 )-\mathcal{U}(\Xi), \;
\mathcal{U}(\Xi)=U_0 \cdot \sinh^2(\pi \Xi /a)
\end{equation}
The Hamilton equations for \eqref{Xi} have an obvious integral of
motion ${H}_E=\mathrm{const}$; boundary conditions give ${H}_E=0$.
Thus, for an instanton solution, we derive $\sqrt{T_0}\sin (\pi
P/P_0 )= \pm \sqrt{\mathcal{U}(\Xi)}$. A simple analysis shows that
this problem has an instanton solution with $\Xi (\pm \infty )=0$
(i.e., $X(\pm \infty )=0$), while the values of momentum differ: $P$
at $\tau\rightarrow -\infty$ and $P+P_0$ at $ \tau \to + \infty $.
This instanton solution describes the tunneling of the kink
chirality.

\begin{figure}[tb]
\includegraphics[width=\figwidth]{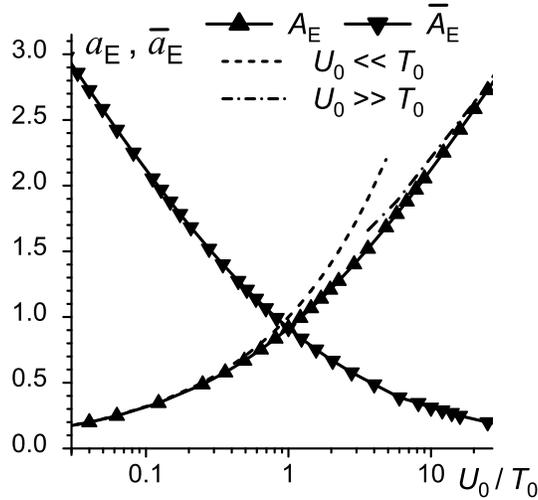}
\caption{\label{AeLog} Dependence on the parameters   $a_E$   and
$\overline{a}_E$, determining the value of the Euclidean action
for the tunneling processes, as a function of $U_0/T_0$ (in
logarithmic scale). Two analytic asymptotic dependencies are also
indicated in the figure with a dashed and a dot-dashed curves.}
\end{figure}

Analysis of the second type of tunneling (tunneling of the kink
coordinate) can be done in a similar manner,  by using a Wick
rotation, substituting $P \rightarrow i\Pi $, and keeping the
coordinate $X$ as a real variable. Then, again, the real Hamilton
function $\bar {H}_E $ for two real variables $X, \;\Pi $ appears,
\begin{equation}
\label{Pi} \bar H_E =U_0 \cdot \sin ^2(\pi X/a)-T_0 \cdot \sinh^2
(\pi \Pi /P_0).
\end{equation}

It is easy to show that this problem has an instanton solution
with $\Pi (\pm\infty)= 0$, $X(\tau \to -\infty)- X(\tau \to
\infty)=a$. This instanton solution describes the tunneling of the
kink from one lattice site to a neighboring one.

For the instanton solutions of both types $\bar H_E,\; H_E =
\mathrm{const} = 0$ the Euclidean Lagrangian reduces to
$L_E=P(dX/d\tau)$. Therefore for both cases, \eqref{Xi} and
\eqref{Pi}, the value of the Euclidean action can be represented as
simple integrals.  For example, for the tunneling of the chirality $
A_E^{0}=\int _0^{P_0}\Xi (P)\;dP$, where
$$\Xi (P)=(a/\pi)\;\mathrm{arcsinh}[\sqrt{(T_0/U_0)}\sin(\pi P/P_0)],$$
resulting in
\begin{equation}
\label{AePi}   A_E^{0}=\left(\frac{2\hbar S}{\pi}\right)a_E
\left(\frac{U_0}{T_0}\right),
\end{equation}
where   $a_E (U_0 /T_0 )$ is a universal function (see
Fig.~\ref{AeLog}) that only depends on the ratio $U_0 /T_0 $. We
can easily find the asymptotic behavior of $a_E $: the functions
$a(z) = \sqrt{z}$ and $a(z) = \ln{z}$, respectively, where
$z=U_0/T_0$. These are shown in Fig.~9.

For the analysis of the tunneling of the coordinate $X$  it is not
even necessary to calculate the corresponding integral $ \bar
A_E^{0}=\int _0^{a}\Pi (X)\;d\!X$. The Euclidean action $\bar {A}_E
$, describing the tunneling of the kink coordinate, is obtained from
the previous formula for the tunneling of the chirality $A_E $ in
(\ref{AePi}) by the replacement $U_0 /T_0 \to T_0 /U_0 $.

The quantity $A_E$, as well as the tunneling splitting  $\Delta
_0$, is a standard measure of the intensity of the quantum
tunneling processes; but there is a physical difference among the
tunnel processes considered here. This is because all of the
values of the kink coordinate of the form $X=an$ (the kink is
found at points of potential energy minima) correspond to
different states of the kink, while all the values of the momentum
differing by $2P_0 $ correspond to the same state of the kink.
Therefore, the processes of tunneling either coordinate or
momentum lead to different physical consequences. \emph{Tunneling
of the coordinate is responsible for the formation of an energy
band in which the number of states with different values of the
quasi-momentum coincides with the number of spins in the chain.}
The width of the corresponding energy band is given in terms of
the tunneling splitting $\Delta _0$, $\Delta E_0 =2\Delta _0$.
\emph{Tunneling of the momentum leads to lifting off the twofold
degeneracy of the states of the kink and causes a splitting of the
level into two, with $\Delta E=2\bar {\Delta }_0$}

\section{Concluding remarks and result discussion}

In conclusion, using  both semiclassical instanton and
perturbative approaches, we study  quantum tunneling effects in
$P-$space and $X-$space  for domain walls in ferromagnetic chains.
We also investigated the quantum dynamics of domain walls
(kink-type solitons) in spin chains. Explicit results have been
obtained for the biaxial model with isotropic interaction $J$, and
rhombic anisotropy with two constants $K_1$ and $K_2$. The
combinations of the two spin interaction constants, $K_1/J$ and
$(K_2-K_1)/K_1$, define two parameters, $U_0$ and $T_0$, for the
effective Hamiltonian describing the quantum dynamics of a domain
wall.

In both limits (small and large $U_0/T_0$) only one type of
transition becomes important, but for the case $U_0 \sim T_0$ the
probability for both transitions are comparable. In this case $U_0
\sim T_0$, the function $a(U_0/T_0)$ is of order of unity, and for
$S \sim 1$ both transition amplitudes are not small; i.e., the
kinks in ferromagnetic chains with spin $S\sim 1$ are essentially
quantum objects. They are characterized by a quantum dispersion
relation (spectrum) of the kink $E=E(P)$, with the presence of
some discrete variable, chirality $\chi = \pm 1$. The quantum
properties of domain walls in a chain with either integer or
half-integer spin are essentially different. For a chain with
integer spin, there are $S$ main energy bands, each one of them
split in two subbands, with a total hybridization of the
chirality. In contrast, for the case of half-integer spin chain we
arrive at a pattern of $2S$ nonoverlapping energy bands, with
chirality hybridization only at some particular points.

Having in mind the case of mesoscopic chain-like artificial
ferromagnetic structures, we discuss the behavior of domain walls
for large  spins $S\gg 1$. At a first glance, for such systems,
the quantum effects should be suppressed by the large spin values.
However, as we have shown, tunneling effects can occurs, even for
values like $S \sim 10^2 - 10^3$, for essentially different values
of the parameters $U_0 $ and $T_0 $. Here the value of the
tunneling exponent can be acceptable $(A_E/\hbar) \leq 15$ - $20$
if $a_E \ll 1$.  We stress the agreement between the probability
of tunneling for one mesoscopic magnetic particle, see Refs.~
\onlinecite{QTM95,MQT_chudn} and the probability of tunneling
processes for the kinks found here.

For domain wall tunneling, at least one of two quantum tunneling
transitions (either tunneling of coordinates or tunneling of
chirality) are  possible. It is useful to introduce the following
empiric rule. The probability of chirality tunneling, including
flipping a large number of spins $N_{\mathrm{kink}} \gg 1$ has the
same order of magnitude as for coherent spin tunneling of a single
particle.

\section*{Acknowledgments}

We gratefully acknowledge partial support from the National
Security Agency (NSA), Laboratory Physical Science (LPS), Army
Research Office (ARO), National Science Foundation (NSF) grant No.
EIA-0130383, JSPS-RFBR 06-02-91200, and Core-to-Core (CTC) program
supported by Japan Society for Promotion of Science (JSPS). SS
acknowledges support from the Ministry of Science, Culture and
Sport of Japan via the Grant-in Aid for Young Scientists No
18740224, the EPSRC via No. EP/D072581/1, EP/F005482/1, and ESF
network-programme ``Arrays of Quantum Dots and Josephson
Junctions''.

\end{document}